\documentclass[twocolumn,trackchanges]{aastex701}

\newcommand{\Oion}{$\rm{O}_{32}$}
\newcommand{\Othree}{[O\textsc{iii}]}
\newcommand{\Otwo}{[O\textsc{ii}]}
\newcommand{\ha}{H$\alpha$}
\newcommand{\hi}{H\textsc{i}}
\newcommand{\hb}{H$\beta$}

\begin{document}

\title{Galaxy mergers drive enhancements in ionization states}

\author[orcid=0000-0003-1767-6421,sname='Le Reste']{Alexandra Le Reste}
\affiliation{Minnesota Institute for Astrophysics, University of Minnesota, 116 Church Street SE, Minneapolis, MN 55455, USA.}
\email[]{alereste@umn.edu}  

\author[orcid=0000-0002-6016-300X]{Kameswara Bharadwaj Mantha} 
\affiliation{Minnesota Institute for Astrophysics, University of Minnesota, 116 Church Street SE, Minneapolis, MN 55455, USA.}
\affiliation{Department of Physics \& Astronomy, University of Missouri-Kansas City, 5110 Rockhill Road, Kansas City, MO 64110, USA.}
\affiliation{Missouri Institute for Defense \& Energy, 5110 Rockhill Road, Kansas City, MO 64110, USA.}
\email{km4n6@umkc.edu}

\begin{abstract}
Recent studies have suggested that galaxy mergers may help drive the escape of ionizing photons from galaxies. However, the ionization properties of merging galaxies have not yet been systematically studied across the full merger sequence in a large, well-defined sample. Here, we investigate the impact of mergers on the line ratio \Oion\ tracing the ionization state and study its evolution as a function of merger progression. We identify 7641 galaxy mergers with robust ($\rm{SNR}>3$) emission line fluxes from the SDSS DR 17, using spectroscopic pair analysis and visual classifications from the participatory science experiment \textit{``Cosmic Disco: Characterizing Galaxy Collisions''}. We find a continuous increase in \Oion\ as a function of merger progression traced by pair separation and visual merger stages, with median and 90th percentile peaking post-coalescence. \Oion\ is significantly enhanced in mergers past the first pericenter passage as compared to a control sample of isolated galaxies. We investigate the properties of galaxy mergers with $\rm{O}_{32}>3$ and find that these are mainly low-mass ($\widetilde{\rm{M}_{*}}=6\times 10^{8}\,\rm{M}{_\odot}$) mergers near coalescence. In this population, large \Oion\ values are due to mergers driving a simultaneous increase in specific star formation rate  and decrease in metallicity. Thus, galaxy mergers enhance the ionization state of galaxies and likely facilitate the escape of ionizing radiation from galaxies, even at low stellar masses. Finally, we infer that Green Pea galaxies, a population characterized by high \Oion, could possibly be low-mass galaxy mergers at or close-to coalescence, a scenario that would reconcile findings from environmental and interferometric 21cm studies.
\end{abstract}


\section{Introduction} 
Galaxy mergers play a crucial role in driving key galaxy evolution processes such as stellar-mass assembly and growth \citep{bell06b,mcintosh_2008,van_der_wel09}, modulation and enhancement of the star formation rates \citep[SFR;][]{Jogee09,Patton11, Ellison13, shah2022}, Active Galactic Nuclei activity \citep[AGN;][]{Treister12,weston17,Hewlett17,shah2020, Ellison2025,LaMarca2025}, and metallicity dilution \citep{Ellison13, thorp2019, Kewley2006a, Rich2012}. Additionally, the role of mergers in driving these changes evolves as a function of the timescale of a merger interaction, with post-coalescence galaxies typically showing the largest change in properties \citep{Ellison2013,Ferreira2025,Ellison2025}. 

Galaxies emitting ionizing radiation (Lyman Continuum, or LyC) drove the key cosmological period known as the Epoch of Reionization \citep{Robertson2010}. Yet, the mechanisms enabling LyC escape from galaxies, necessary to ionize the intergalactic medium, remain poorly known. Because galaxy mergers trigger starbursts and disrupt the neutral gas that otherwise impedes LyC emission, they were proposed early on as potential mechanisms for LyC escape \citep{Bridge2010,Bergvall2013}. While recent simulations and observational studies have found indications that mergers do indeed facilitate LyC escape \citep[][although see \citealp{Mascia2025}]{LeReste2024,Maulick2024,Kostyuk2025,Zhu2025,LeReste2026,Ejdetjarn2026,Rivera-Thorsen2026}, no study has yet measured LyC emission across a statistically large number of galaxy mergers sampling the full merger sequence. As a result, there is currently no consensus on whether mergers generally facilitate LyC escape, if they do play a role in LyC escape from low-mass galaxies, nor on what merger configurations are conducive to this escape. Previous investigations were limited by small samples, complicated selection functions, and observational bias. 
These limitations persist because direct LyC detection is inherently challenging. At lower redshifts ($z < 1$) where merger identification is most accurate \citep[see][]{mantha2019studying}, the low throughput of UV observations requires large time investments to yield a few tens of detections \citep[e.g.][]{Flury2022a}. At higher redshifts ($z > 2$) LyC is redshifted to the optical range, allowing more efficient observations. However, galaxy mergers become increasingly difficult to detect at higher redshift due to cosmological  surface brightness dimming \citep{mantha2019studying}. Additionally, the opacity of the intergalactic medium  increasing with redshift \citep{Inoue2014} severely biases LyC observations.

\begin{figure}
    \centering
    \includegraphics[width=0.80\linewidth]{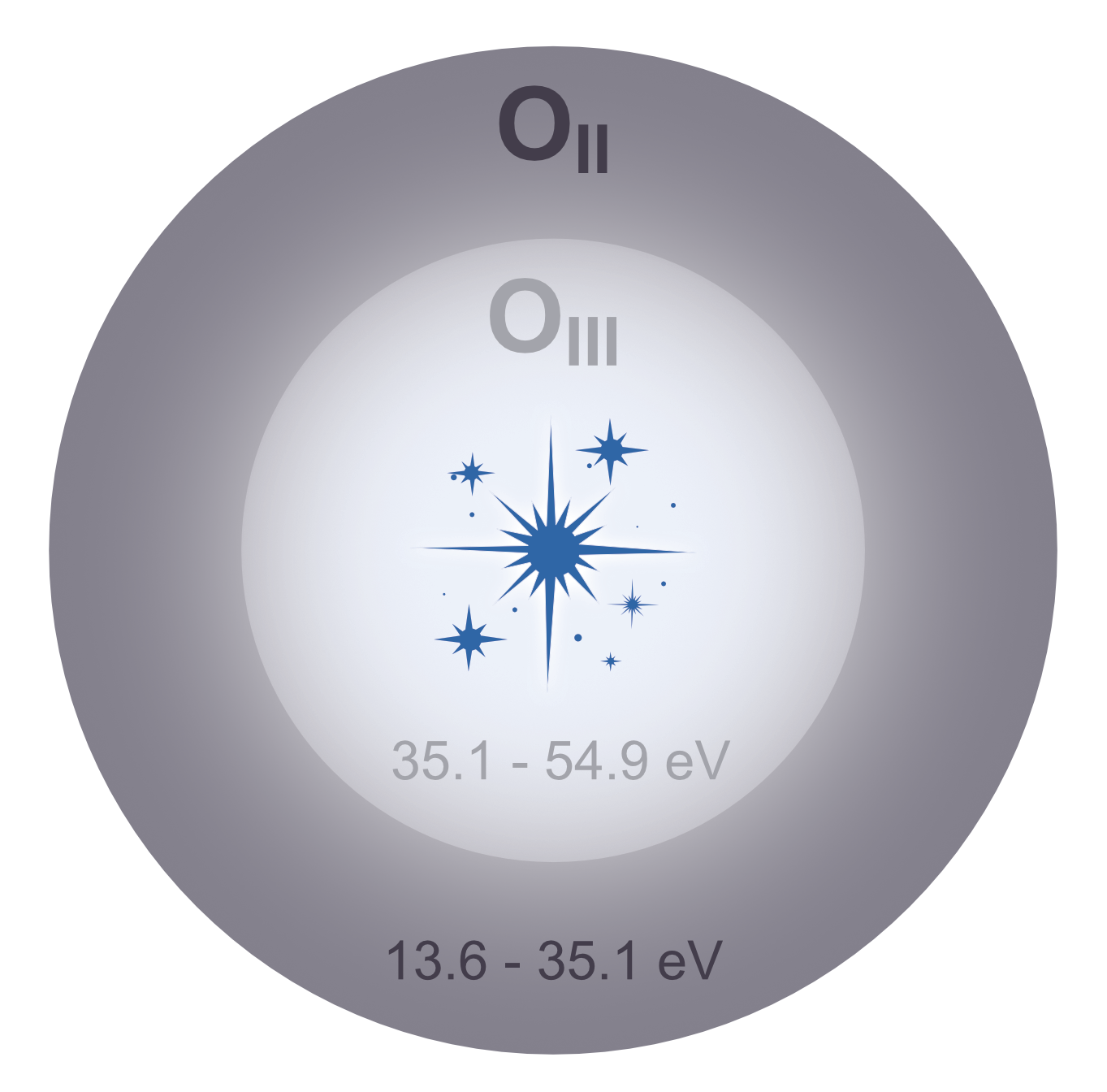}
    \caption{ Schematic illustration of oxygen ionization stratification in an idealized H\textsc{ii} region, demonstrating how $\rm{O}_{32} \equiv [\rm{O}\textsc{iii}] / [\rm{O}\textsc{ii}]$ probes the ionization state. The concentric layers correspond to zones defined by the ionization potentials of oxygen (O\textsc{i} $\rightarrow$ O\textsc{ii}: $13.6\,$eV, O\textsc{ii} $\rightarrow$ O\textsc{iii}: $35.1\,$eV, O\textsc{iii} $\rightarrow$ O\textsc{iv}: $54.9\,$eV). Singly ionized oxygen (O\textsc{ii}) is found in environments filled with photons with $h\nu > 13.6\,\rm{eV}$, but it is depleted via further ionization into doubly ionized oxygen (O\textsc{iii}) where photon energies exceed $h\nu>35.1\,\rm{eV}$. At fixed ionizing spectrum hardness, gas density, temperature, and metallicity, \Oion\ characterizes the relative dominance of high- and low-ionization zones, serving as an observational tracer of the interstellar medium ionization state.}
    \label{fig:O32_illustr}
\end{figure}

Given the strong observational challenges associated with direct LyC detection, secondary tracers calibrated at low-redshift are often employed to study LyC escape \citep{Jaskot2025}. 
 The oxygen line ratio \Oion=\Othree/\Otwo\, is an important observable that characterizes the ionization state of a galaxy \citep[][and see illustration in Figure \ref{fig:O32_illustr}]{Kewley2019}. This line ratio has gained significant attention in LyC studies in the past decades, as high \Oion\ values have often successfully been used to identify the rare galaxies emitting ionizing radiation in the local Universe \citep[e.g.][]{Izotov2018,Flury2022a}. Additionally, \Oion\ and the ionizing photon escape fraction are weakly but significantly correlated \citep[][]{Izotov2018,Chisholm2022,Flury2022b}, making it an important indicator of LyC escape. It is generally agreed upon that a high \Oion\ value is a necessary condition for LyC to escape a galaxy \citep{Nakajima2020}.
 
Here, we present a characterization of \Oion\ in galaxy mergers selected from the Sloan Digital Sky Survey Data Release 17 \citep[SDSS DR17 ;][]{Smee2013,Blanton2017,Abdurrouf2022}. Galaxy mergers are identified through spectroscopic pair analysis and visual identification to fully sample the breadth of the merger sequence. Visual identification and characterization of merger stages is provided by Zooniverse volunteers through a participatory science project called \textit{``Cosmic Disco: Characterizing galaxy collisions''}. With this sample, we seek to answer three questions: i) whether galaxy mergers have different \Oion\ values than  control isolated galaxies, ii) if any trends between merger progression and \Oion\ can be found, and iii) what the properties of high-ionization state mergers are. We describe the sample and methods for merger analysis in \ref{sec:methods}. Results are presented in \ref{sec:results}, and their implications for LyC emission, and for the nature of galaxy populations with high \Oion\ are discussed in \ref{sec:discussion}. Finally, we present a short conclusion in \ref{sec:conclusion}.

In the following, we use a standard flat $\Lambda$CDM cosmology with $H_0 = 70\, \rm{km}\,\rm{s}^{-1}\,\rm{Mpc}^{-1}$ and $\Omega_m=0.3$.

\section{Methods} \label{sec:methods} 
The primary goal of this work is to assess the trends in \Oion, which we measure here as \Othree$_{\lambda 5007}/$ \Otwo$_{\lambda 3727}$ as a function of merger stage, in comparison with non-merging control galaxies. To achieve this, we perform detailed close-pair analysis and participatory science-driven merger characterization on a sample of stellar-mass complete, star-forming galaxies spanning $0.01\leq z \leq 0.055$ with significant emission line detections in SDSS DR17. In this section, we provide detailed descriptions of our core sample selection process and subsequent quality control steps, and describe both the close-pair analysis and participatory science experiment design.

\subsection{Sample selection}
We obtain a general emission line galaxy sample by selecting targets from SDSS DR17 with MPA-JHU emission line and stellar mass measurements available \citep{Brinchmann2004,Kauffmann2003a}. Since \Oion\ is strongly anti-correlated with $\rm{M_*}$, we include galaxies with stellar masses down to $\rm{M_*}>10^8\,\rm{M_\odot}$. This threshold both ensures that galaxies with high \Oion\ are included in the sample and allows the investigation of the impact of mergers on the ionization properties of low-mass galaxies. To ensure completeness over this mass range,
we select galaxies with redshift $z=0.01$ to $z=0.055$. We then apply signal-to-noise cuts, requiring $\rm{SNR}>3$ in \Othree$_{\lambda 5007}$, \Otwo$_{\lambda 3727}$, \ha, \hb, and [NII]$_{\lambda6583}$. These lines are necessary to calculate \Oion, estimate the nebular dust extinction and exclude AGNs from the sample. MPA-JHU line fluxes are already corrected for Milky Way extinction using the extinction curves in \cite{ODonnell1994}. We additionally correct line fluxes for internal dust extinction using the observed Balmer decrement and \cite{Cardelli1989} dust extinction law, assuming $R_v=3.1$ and intrinsic \ha/\hb$=2.87$. AGNs are excluded using the Baldwin Phillips Terlevich \citep{Baldwin1981} diagram-based selection criterion outlined in \cite{Kauffmann2003b}. Finally, we remove galaxies with anomalous r-band correction factors (\texttt{spectofiber}) leading to excessive line fluxes by only retaining galaxies with  $f_{H\alpha}\leq1\times10^{-12}\, \rm{erg}\,\rm{s}^{-1}\,\rm{cm}^{-2}$ \citep{Lee2026}. This leads to a final sample of 38805 emission line galaxies. We calculate SFRs using the \ha\ emission line fluxes and the \cite{Kennicutt1998} relation. Additionally, we estimate oxygen abundances using the [N\textsc{ii}]$_{\lambda6583}$ and \ha\ line fluxes and the N2 calibration presented in \cite{Marino2013}.

\subsection{Spectroscopic close-pair identification}
 To assess the change in \Oion\ over a range of merger stages, we first identify galaxies involved in a spectroscopic close pair. Following standard approaches in the literature, we identify spectroscopic pairs using cuts in projected physical separation ($D_{\rm sep}$) and line-of-sight velocity difference ($\Delta v$). Various studies that assessed the role of mergers in star-formation rate and AGN activity enhancements \citep[e.g.][]{Patton2013, shah2020, shah2022} have adopted projected separation distances $D_{\rm sep}$ spanning tens to $ 100 - 150\,{\rm kpc}$ and $\Delta v \sim 500 - 1000\,{\rm km/s}$. These cuts are motivated by simulation-based predictions that these cuts isolate likely bound systems and merger candidates on timescales $\sim 0.5-1.5\,{\rm Gyr}$ \citep{lotz2010, lotz2011,Ventou2019}. Inspired by the selection cuts in these prior studies, we adopt an inclusive spectroscopic pair-selection criteria $D_{\rm sep}\leq 150\,{\rm kpc}$ and $\Delta v \leq 1000\,{\rm km/s}$. This broad selection window captures galaxies experiencing potential early-stage tidal influence. The projected separation can subsequently be used as a continuous or binned proxy for interaction stage, serving as a complementary axis to the visually classified merger stages described below. Finally, we limit the sample to pairs that have comparable mass ratios by selecting $M_{1}/M_{2} \leq 1/4$ \citep[often referred to as ``major mergers''; see][]{mantha2018}, identifying $4735$ spectroscopically confirmed pairs. 

\subsection{Cosmic Disco: Visual Characterization of Merger Stages}

From previous studies, we expect galaxy mergers close to nuclear coalescence to have the largest LyC escape fraction and ionization state due to the enhanced SFR observed in late-stage mergers \citep{Kostyuk2025,Ejdetjarn2026,LeReste2026}.  
However, while spectroscopic close-pair selection provides a direct way to identify galaxies with nearby companions in potential physical proximity, projected separation alone is an incomplete tracer of merger stage. Galaxies at different times along an interaction can have similar projected separations, and late-stage or post-coalescence mergers may no longer contain two clearly separable nuclei with independent spectra. To better characterize the \Oion\ trend across the full merger sequence, we complement our spectroscopic pair identification procedure with visual characterizations from a dedicated participatory science project: \textit{``Cosmic Disco: Characterizing Galaxy Collisions''}\footnote{\url{https://www.zooniverse.org/projects/kbmantha/cosmic-disco-characterizing-galaxy-collisions}} (hereafter referred to as Cosmic Disco). In this project, volunteers were tasked to identify and classify mergers into one of four merger stages. We describe the project in \ref{sec:CD-methods}.

\subsubsection{Project Data Preparation}\label{sec:CD-methods}
The goal of Cosmic Disco is to provide a general catalog of galaxy mergers with visual merger stage classification. We first start from a broad sample by selecting all galaxies from SDSS DR17 with spectroscopic information. We cross-match this sample with the DESI Legacy Imaging Surveys \citep{Dey2019} to obtain images for volunteers to classify. DESI  has considerable overlap with SDSS but provides deeper imaging, better suited to the identification of the faint tidal features characterizing late-stage mergers. 

To help focus participatory science efforts on potential mergers, we cross-matched our target galaxies with the Zoobot morphological catalog \citep{Walmsley2022}, using a matching radius of 2\arcsec. Zoobot is a deep-learning model trained on a vast collection of images to emulate the visual morphological classifications of the Galaxy Zoo project. It has been successfully applied to $\sim 8.7\,{\rm M}$  galaxies in DESI, and these Galaxy Zoo-like classifications are tabulated for public use \citep{WalmsleyZenodo}. For each galaxy, the Zoobot table includes probabilistic scores for various morphological categories, including galaxy mergers. Using the cross-matched SDSS/DESI/Zoobot catalog, we used the {\tt legacystamps} pipeline\footnote{\url{https://github.com/tikk3r/legacystamps}} to download three-color RGB images ($grz$) of galaxies with spectra to be visually classified by the volunteers. Because mergers and tidal features can have  spatial extents spanning several kiloparsecs, we downloaded each image in two configurations to facilitate flexible visual inspection. These provide a zoomed-in and a zoomed-out view, spanning $75\,{\rm kpc}$ and $150\,{\rm kpc}$ on a side, respectively. This choice for image sizes was motivated by prior close-pair studies, which identify $\sim 150\, {\rm kpc}$ as the maximum projected separation at which galaxy pairs significantly impact star-formation rates \citep{Patton2013}. 

For Cosmic Disco, to focus the visual merger stage characterization efforts on systems that are likely to be merging candidates, we select galaxies that have \texttt{merging\_merger\_fraction} $>0.1$ from Zoobot. While this fraction may seem low, participatory science experiments have shown early-on that merger vote fractions are generally lower than expected, even for clear mergers \citep{Lintott2008,Darg2010}. Since Zoobot emulates human classifications, a low threshold is the best way to securely include mergers across the sequence, including late-stage mergers with faint tidal features. We select this threshold by performing visual inspection of a randomly selected collection of images in bins of \texttt{merging\_merger\_fraction} across its entire range, finding that \texttt{merging\_merger\_fraction} $< 0.1$ is when most images resemble non-merging/morphologically unperturbed systems. This cut yields a sample of $7244$ galaxies to be classified by volunteers.

\subsubsection{Project Task Setup \& Outcomes}
 With our primary merger candidate sample of $7244$ galaxies, we created a dedicated workflow within Cosmic Disco where volunteers were tasked with characterizing each presented image into one of the following categories:

\begin{itemize}
    \item {\it Pre-Interaction Stage}: A pre-interaction stage is when two (or more) galaxies are in the nearby vicinity but {\bf do not} have any disturbed morphological features.
    \item {\it Post-interaction Stage}: If two (or more) galaxies are distinctly separated from each other and either {\bf one} or {\bf both} display disturbed morphological features.
    \item {\it Near Coalescence}: If two (or more) galaxies that are {\bf very} close (on the order of a single galaxy’s size) or embedded within each-other’s light envelopes displaying disturbed morphological features
    \item {\it Post Coalescence}: If predominantly one central galaxy core region is observed with strong indications of disturbed morphological features.
    \item {\it Not-a-Merger}: A single galaxy with no other galaxy in its vicinity and {\bf not} hosting any disturbed morphological signatures.
    \item {\it Artifact/Bad Image}: Owing to any observational errors or missing band coverage, some images may come with poor visual quality making them unfit for robust characterization.
\end{itemize}

 Following recommendations for Zooniverse project setup and drawing inspiration from prior Galaxy Zoo projects, we created a tutorial explaining the key task structure and walkthrough of an example classification. Additionally, we provided a detailed visual ``field guide'' in which several curated image examples for each of the six categories above are provided along with an expert written reasoning on why they belong to that certain category. To minimize the variation from individual volunteer choices, we require that every galaxy image be classified by $20$ volunteers before it is taken out of circulation (referred to as ``retirement''). We launched our project on January 9th, 2025 and over the course of three months, $\sim1100$ registered volunteers provided $\sim 145,000$ individual classifications. We export the raw classification data from Zooniverse's project builder's infrastructure to generate consensus answers for each galaxy and use it for downstream scientific analysis.

\begin{figure*}[t]
    \centering
    \includegraphics[width=\textwidth]{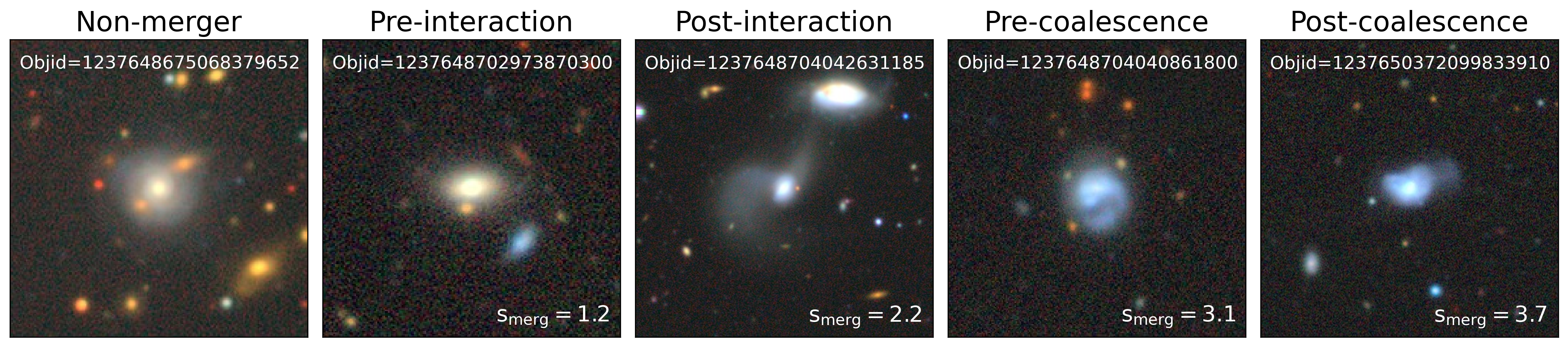}
    \caption{Example Cosmic Disco galaxies for the different merger stage classifications. From left to right, the panels show DESI Legacy Imaging Survey RGB ($grz$) color-composite images for a non-merger, a pre-interaction merger, a post-interaction merger, a pre-coalescence merger, and a post-coalescence merger. SDSS object identifiers are shown on the top of each panel, and for merging systems the merger stage is indicated on the bottom right corner. }
    \label{fig:thumbnails}
\end{figure*}

Using the raw classification answers for each galaxy, we compute a merger stage score $s_{\rm merg}$. We do this by numerically encoding each merger category between $1$ (pre-interaction) and $4$ (post-coalescence) and compute the average score for each galaxy over the merger-specific category votes. 

\begin{equation}
    s_{\rm merg} = \frac{\Sigma v_{i}}{N},
\end{equation}

where $v_{i}$ is the volunteer-chosen merger category converted into an integer ($1-4$) and $N$ is the total number of votes toward any merger category. In Figure\,\ref{fig:thumbnails}, we show example galaxies as per their $s_{\rm merg}$ values. In subsequent analysis, we use $s_{\rm merg}$ alongside the projected separation as tracers of merger progression. 

For the \Oion\ analysis carried out here, we remove Cosmic Disco galaxies that have been voted as having artifacts by at least half of the volunteers.
We apply the same selection cuts to the Cosmic Disco and pair samples as those applied to the full emission line catalog (SNR$>3$, AGN removed, galaxies with anomalous emission line fluxes excluded).
Finally, to only select secure galaxy mergers, we additionally only retain Cosmic Disco galaxies that have been voted by at least half of the volunteers to be in any merger category. 
This yields a catalog of 2927 visually-identified mergers and 4735 galaxies in spectroscopic pairs with valid, high signal-to-noise-ratio emission lines fluxes where \Oion\ can be analyzed.

\subsection{Isolated Control Galaxy Selection \& Matching}

In this work, we assess the impact of galaxy mergers as a function of their merger stage by comparing their ionization state (measured via \Oion) to isolated control galaxies. We select a pool of non-merging, potential control galaxy sample by choosing a subset of galaxies in the emission line catalog with low Zoobot merger probability $f_{merger}<0.05$ that do not have any spectroscopically-confirmed neighbors within $D_{\rm sep}\leq 150\,{\rm kpc}$. In total, we find $27501$ isolated galaxies to be used as possible controls.

To control for systematics during the analysis of \Oion\ in galaxy mergers, we search for the ``best matching'' control galaxy according to their properties. These include the redshift, stellar mass, and local environment. We use the number of companions within a $2\,{\rm Mpc}$ projected distance ($N_{2}$) as a tracer of local environment, as it is commonly used in the literature as a proxy for galactic environmental neighborhood \citep[see][]{patton2016}. We do not control for the star formation rate or metallicity as traced by the oxygen abundance $12+\rm{log(O/H)}$, as those properties are known to significantly differ in mergers as compared to isolated galaxies \citep{Patton13,Bustamante2018}. We require that a control satisfies $\Delta z<0.01$, $\Delta \rm{log}M_{*}<0.2$ and $\Delta N_{2}<2$ for mergers with $N_{2}\leq20$, or  $\Delta N_{2}< 0.1\,N_{2}$ otherwise. To match each merger to the control galaxy closest in properties, 
we compute the euclidean distance across $z$, $\rm{log}\,M_{\star}$ and $N_{2}$, and select the control galaxy as that minimizing the distance. The same isolated galaxy can act as a control to multiple galaxy mergers. Five galaxies do not have controls, likely due to their high $N_{2}\geq25$ ; we exclude those from the sample. {In Appendix Figure \ref{fig:controlvmerg}, we show distributions for the difference of values between mergers and their controls for the redshift, stellar mass and N2. The distributions are centered around zero, indicating the matching is adequate.}

\section{Results}\label{sec:results}
Here, we present results regarding the \Oion\ properties of galaxy mergers as compared to matched isolated galaxies. In particular, we compare the evolution of \Oion\ with merger progression to values expected in isolated controls and assess what causes elevated \Oion\ values in galaxy mergers. 

\begin{figure*}[t]
    \centering
    \includegraphics[width=\textwidth]{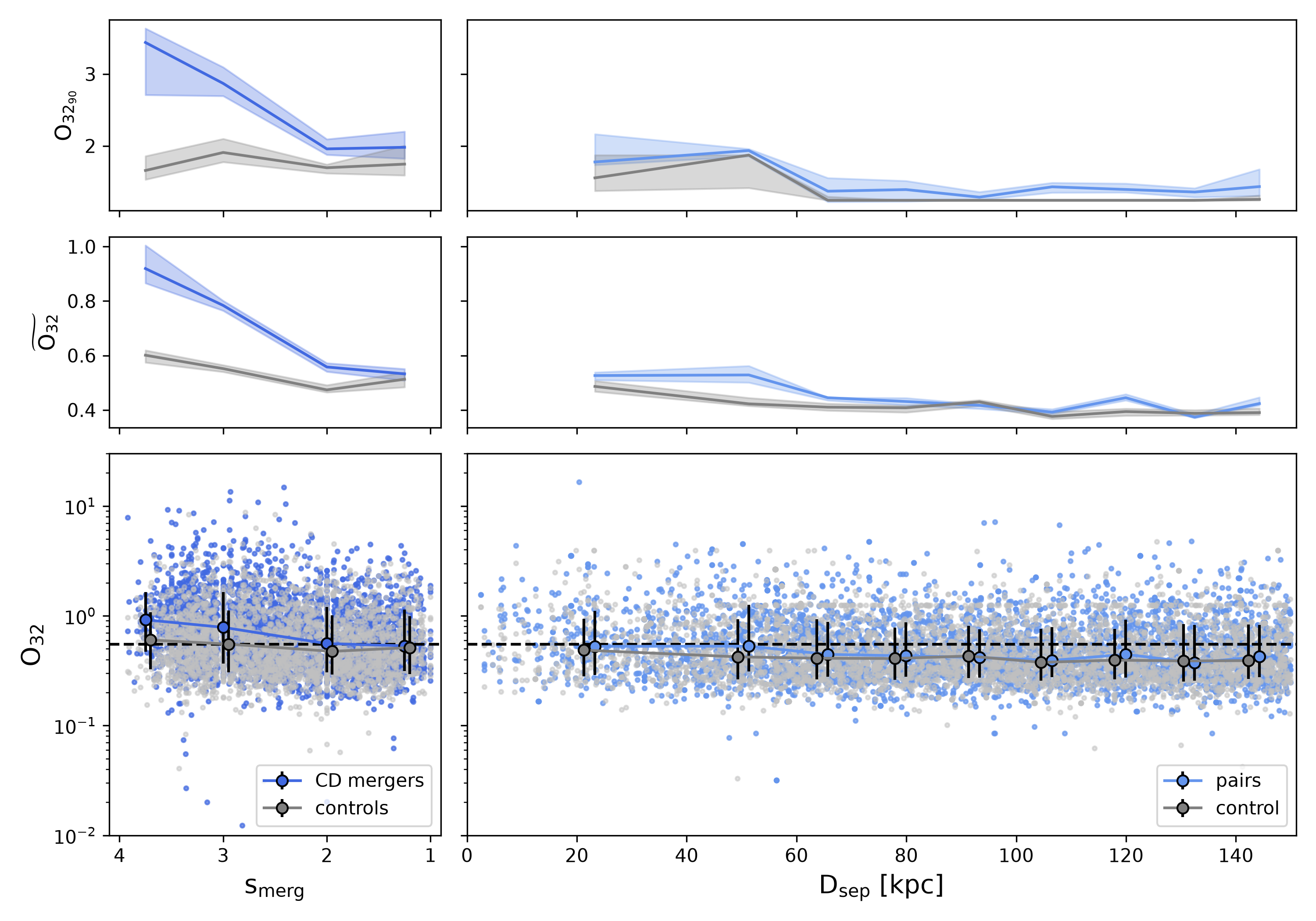}
    \caption{Evolution of \Oion\ as a function of merger advancement. \textbf{Left:} \Oion\ evolution across visually-identified merger stages ordered from early stage to post-coalescence from right to left. \textbf{Right:} \Oion\ evolution as a function of projected pair separation. The rows display, from top to bottom, the 90th percentile $\rm{O}_{32_{90}}$, the median $\widetilde{\rm{O}_{32}}$, and \Oion\ for individual galaxies with binned medians and bars indicating the 16th and 84th percentile. In all panels, mergers are indicated by blue data points and curves, while gray data points and lines represent control galaxies matched in redshift, stellar mass and local environment. Shaded regions indicate the $1\sigma$ confidence interval obtained from bootstrapping. The black dashed line shows the median $\text{O}_{32}$ for the full emission line catalog. A strong enhancement in $\text{O}_{32}$ is observed in the merger sample relative to the controls at advanced merger stages.}
    \label{fig:O32_evol}
\end{figure*}
\subsection{Do mergers have different $O_{32}$, and how does $O_{32}$ evolve with merger stage?}
Figure \ref{fig:O32_evol} shows the evolution of the ionization state as traced by \Oion\ as a function of the merger progression, where more advanced mergers are found on the left, and incoming pairs on the right. The rightmost panels shows \Oion\ as a function of pair separation $D_{\rm sep}$, while the leftmost panels show visually classified merger stages in Cosmic Disco. Values of \Oion\ for mergers (blue points and curves) are compared to those for control isolated galaxies selected to match the redshift, $\rm{M}_{*}$ and local environment (gray points and curves). We calculate the median $\widetilde{\rm{O}_{32}}$ and 90th percentile $\rm{O}_{32_{90}}$ for the mergers and control galaxies in bins, with uncertainties measured using $10^4$ bootstrap iterations. For pairs, we evaluate $\widetilde{\rm{O}_{32}}$ and  $\rm{O}_{32_{90}}$ in 9 bins with equal number of galaxies. For visually-classified mergers, we define the bins based on $s_{merg}$ ($1-1.5$, $1.5-2.5$, $2.5-3.5$, $3.5-4$).

Both $\widetilde{\rm{O}_{32}}$ and $\rm{O}_{32_{90}}$ increase as pair separation decreases.  We run Kolmogorov-Smirnov (KS),  Anderson Darling (AD) and Mood's median tests to assess whether the \Oion\ distributions and medians differ statistically between mergers and control galaxies (see Appendix Figure \ref{fig:O32_stats}). Despite the apparent increase with merger progression for pairs, p-values remain elevated ($p>1.3\times 10^{-3}$) in all bins,  indicating no significant differences are found between the \Oion\ properties of close pairs and their corresponding control galaxies. For visually-identified galaxy mergers, we observe a strong evolution of \Oion\, with both the median and 90th percentile values sharply increasing from early merger stages 
($\widetilde{\rm{O}_{32}}=0.53$, $\rm{O}_{32_{90}}=1.98$ for $s_{merg}\sim1$, incoming pairs) to post-coalescence mergers ($\widetilde{\rm{O}_{32}}=0.92$, $\rm{O}_{32_{90}}=3.44$ for $s_{merg}\sim4$). The \Oion\ values for control galaxies also increase slightly with increasing $s_{merg}$, however, statistical tests confirm the samples differ from $s_{merg}=1.5$. In particular, in all statistical tests, we find $p<1.3\times 10^{-3}$ for $s_{merg}>1.5$, indicating the null hypotheses that the samples are drawn from the same distribution and that the samples come from distributions with the same median can be rejected with more than $3\sigma$ confidence. In other words, both the median and 90th percentile increase with merger stages for visually-identified mergers, and decouple from the distribution of control galaxies past $s_{merg}=1.5$. Taken together, these results indicate that past the first pericenter passage, visually-identified galaxy mergers have significantly larger \Oion\ values, and count a larger fraction of galaxies with large \Oion\ values than isolated galaxies with otherwise similar properties. The low statistical significance of the difference in \Oion\ between pairs and their controls is likely due to confusion introduced by pair classification, and to the relatively low number of very close pairs ($\rm{D}_{sep}\leq20\,\rm{kpc}$) in our sample. While stellar tidal features clearly indicate that at least one pericenter passage has already occurred, a small projected separation could be observed in distant pairs due to the angle through which the system is viewed. 

\begin{figure*}[t]
    \centering
    \includegraphics[width=\textwidth]{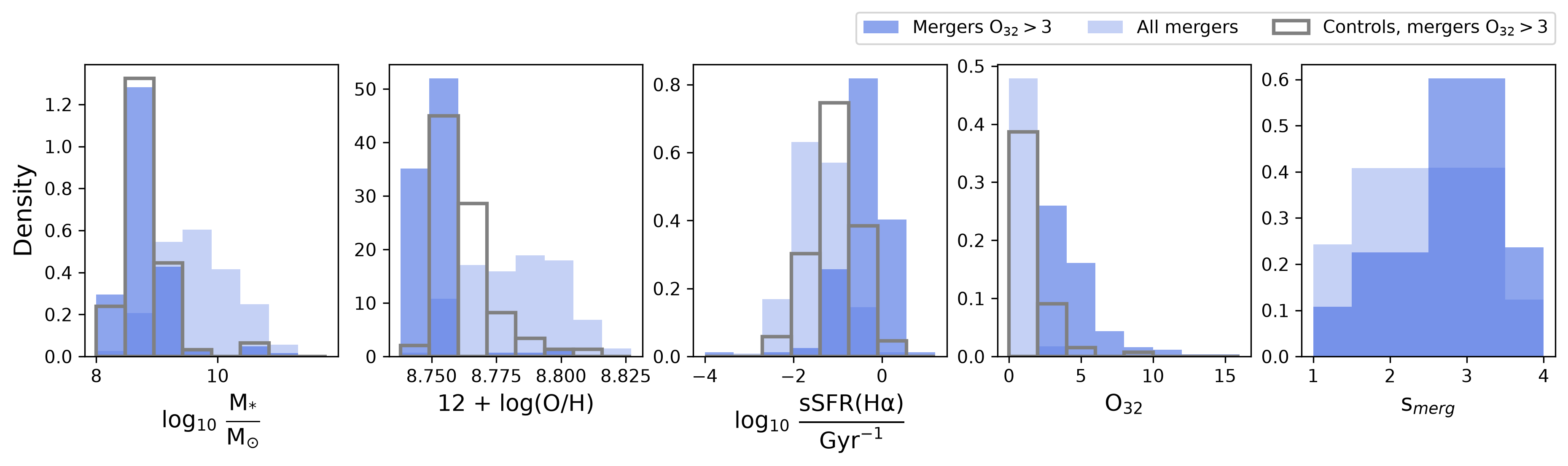}
    \caption{Properties of galaxy mergers with high \Oion\ (\Oion$>3$, dark blue) as compared to those of their matches in the isolated control sample (gray outline) and to the full merger sample (light blue). From left to right, the panels show density plots for the stellar mass, the oxygen abundance, the specific star formation rate calculated from the H$\alpha$ emission line flux, \Oion\ and the merger stage. The highest values of \Oion\ in galaxy mergers are found in low-mass galaxy mergers at late interaction stages, indicating these high \Oion\ values are due to merger-driven decrease in metallicity and increase in specific star formation rates. }
    \label{fig:O32_high}
\end{figure*}

\subsection{What drives high \Oion\ values in galaxy mergers?}
 In Figure \ref{fig:O32_high}, we show the properties of all galaxy mergers (including pairs and visually-classified mergers) with \Oion$>3$, as compared to those of their control galaxies and of the full merger sample. We use this specific threshold for \Oion\ as it was one of the criteria that enabled the selection of galaxies emitting ionizing emission in the largest low-z LyC survey to date \citep[][]{Flury2022a}. The properties shown in Figure \ref{fig:O32_high} are known to impact \Oion; these include the stellar mass, the oxygen abundance, the specific star formation rate (sSFR, defined as $\frac{\rm{SFR}}{\rm{M}{*}}$), and the merger stage. It is apparent that mergers with high \Oion\ are part of a sub-population of galaxy mergers distinct in properties from the parent sample including pairs and visually-identified galaxies. In particular, high \Oion\ mergers have much lower stellar masses than the general sample ($\widetilde{\rm{M_*}}=6.2 \times 10^{8} M_{\odot}$), lower oxygen abundances (median $12+\rm{log}(O/H) = 8.5$), and higher specific star formation rates ($\widetilde{sSFR}=0.5\,\rm{Gyr}$) than are found in the full merger sample. Finally, while  high \Oion\ mergers sample a range of merger stages, they tend to be advanced mergers, close to or post nuclear coalescence ($\widetilde{s_{merg}}=2.9$). 
We note that these results do not change significantly when only considering visually-identified galaxy mergers in Cosmic Disco with high \Oion.

 As expected from the control selection process, galaxy mergers and their controls have similar stellar mass distributions. However, galaxy mergers with high \Oion\ have much lower oxygen abundances than their controls, and higher specific star formation rates. As a merger interaction progresses, the metallicity is expected to decrease while star formation rate increases \citep{Kewley2006a,Ellison13,Faria2025}. A study of Cosmic Disco galaxies found that the specific star formation rate increases as a function of $s_{merg}$ \citep{Lee2026}, so trends expected from results in spectroscopic pairs likely apply here. \Oion\ is anti-correlated with metallicity, and correlated with the specific star formation rate \citep[][]{KewleyDopita2002,NakajimaOuchi2014}. Therefore, the increase in \Oion\ in advanced mergers as compared to isolated control galaxies can be explained by merger interactions simultaneously driving a decrease in metallicity while increasing the specific star formation rate.

A question that naturally emerges from this is whether the merger stages at which we find most galaxies with high O$_{32}$ ($s_{merg}=2.5-3.5$) coincide with the timescales in which mergers are expected to raise the sSFR and dilute the gas-phase metallicity. To test this would require the intrinsic merger timescale distributions of visually-identified merger categories. However, the visual merger stages and the $s_{merg}$ measurement have not yet been calibrated against physical merger timescales. Nonetheless, existing studies suggest  broadly consistent timescale regimes for sSFR enhancement and drop in O/H. In cosmological simulations, SFR enhancements and central metallicity dilution appear in close pairs, before coalescence \citep{Patton13, Bustamante2018}, and SFR enhancements persist for 500\,Myr after coalescence \citep{Byrne-Mamahit23}. Idealized merger simulations similarly show sSFR enhancements around the later pericentric passages, accompanied by changes in the gas-phase O/H gradient driven by strong central gas inflows \citep{Sillero17}. Observationally, post-mergers in MaNGA show enhanced star formation and uniformly suppressed metallicity out to 2 effective radii \citep{thorp2019}.  These merger stages, from the final pericentric passages to a few hundred Myr after coalescence, correspond qualitatively to the stages where we observe the highest O$_{32}$. However, we caution that these studies focus on massive galaxies ($M_* \gtrsim 10^{10}\,M_\odot$), which are more than an order of magnitude above the median stellar mass of our high-O$_{32}$ mergers. The shallower potentials and higher gas fractions of low-mass galaxies may change both the amplitude and duration of merger-induced enhancements and the intrinsic timescales over which they are detected through visual inspection. Calibrating $s_{merg}$ against simulation-derived intrinsic timescales will be needed to test this more directly.
 
\section{Discussion}\label{sec:discussion}
In this section, we briefly discuss the implications of our results. Specifically, we address how it adds to our understanding of the impact of mergers on ionizing radiation, we make inferences on the nature of the high \Oion\ galaxy population known as ``Green Peas'', and discuss the caveats impacting our work.

\subsection{Implications for merger-driven LyC emission}
Observational studies of the role of galaxy mergers on ionizing radiation escape \citep[e.g.][]{Zhu2025,Mascia2025,LeReste2026} so far have performed merger identification on samples of galaxies already observed in LyC. While these studies have shown that some mergers are LyC-emitters, limited conclusions can be drawn from these samples regarding mergers as a general mechanism facilitating LyC emission from galaxies. Here, we have measured \Oion, a tracer of LyC escape, across a general sample of galaxy mergers with robust classifications. In addition to helping better assess the role of mergers in changing ionization states, we were able to identify the merger properties  leading to high ionization states, and thus higher LyC escape probabilities. 

In \cite{LeReste2026}, merger stages were visually characterized in a similar way as here. In this study, LyC-emitting mergers were all identified as post-interaction and near-coalescence mergers, leading the authors to conclude LyC emission may only happen at the end of a merger interaction. In this work, we confirm these initial results : while high \Oion\ galaxies can be found at a range of merger stages, they have merger stages predominantly characteristic of near-coalescence ($s_{merg}=2.5-3.5$). This result therefore adds to the stack of evidence pointing to late-stage mergers as drivers of LyC escape.

Galaxy mergers have so far been inferred to enable LyC escape primarily in galaxies with high stellar masses ($\rm{M}_{*}>10^{9}\,\rm{M_{\odot}}$), where the interaction would be necessary to sufficiently disturb the neutral interstellar medium \citep{LeReste2026}. In a large fraction of low-mass galaxies, star formation alone may be able to disrupt \hi\ though outflows due to shallow gravitational potentials \citep{Cenci2024}, and be sufficient to create the conditions enabling escape. Similar inferences have been made on the role of mergers for Lyman-$\alpha$ emission, a resonant line of \hi\ that traces LyC escape \citep{Hagen2016,LeReste2025a,Ren2026}.
Here, we demonstrate that low-mass galaxy mergers in fact have much higher \Oion\ than their low-mass isolated counterparts. This has two immediate consequences. First, it shows for the first time that galaxy mergers also likely contribute to LyC escape from low-mass galaxies. Second, it has implications for the identification of LyC-emitting galaxy mergers.
 Due to surface brightness dimming proceeding as $(1+z)^4$, the tidal features needed to characterize late-stage mergers are hard to detect at high redshift \citep{mantha2019studying}. This implies that a large number of low-mass, high-z LyC emitters may be incorrectly identified as isolated galaxies. Interestingly, \cite{Mascia2025}, using morphometrics to identify mergers in likely LyC-emitters at $z\sim6$, found a low number of galaxy mergers, and concluded compact isolated galaxies are more likely to be LyC-emitters. This contrasts with results at lower redshift ($z\lesssim3$), finding a large fraction of galaxy mergers ($\sim40-80\%$) in samples of LyC-emitting galaxies \citep{Zhu2025,LeReste2026}. This discrepancy could be interpreted as an evolution of the galaxy properties that drive LyC escape with redshift. However, our identification of low-mass, late-stage galaxy mergers as the merger population with highest \Oion\ suggests high-z results are likely biased by observational limitations. This implication extends well beyond ionization properties. If late-stage mergers have a stronger impact on galaxy properties than early-stage mergers at high-z, as they do in the low-z universe, the impact of mergers on galaxy properties in the early Universe is likely severely underestimated.

\subsection{Green pea galaxies : a population of low-mass galaxy mergers post-coalescence?}

\begin{figure*}[t]
    \centering
    \includegraphics[width=0.95\linewidth]{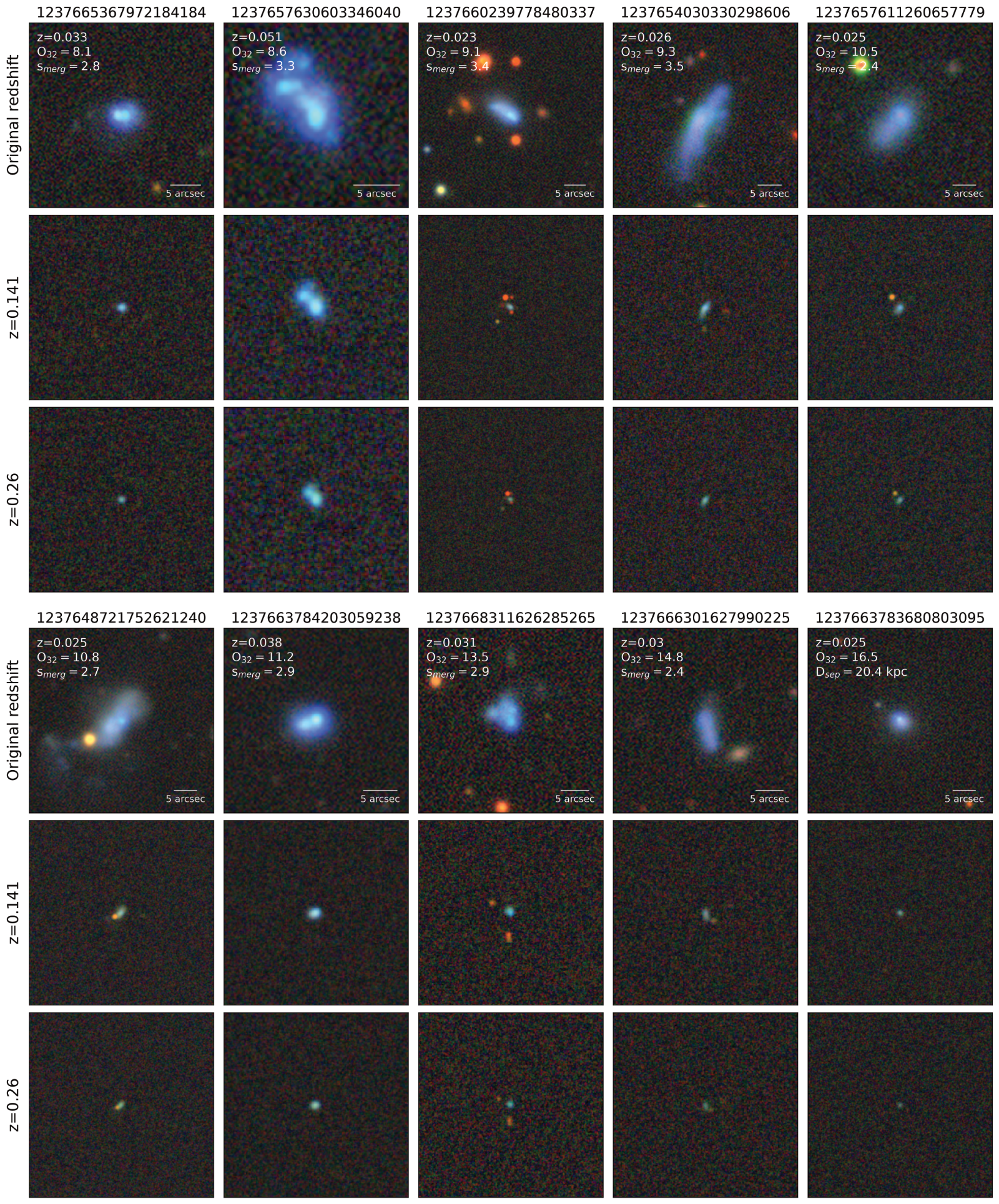}
    \caption{DESI Legacy Imaging Survey RGB ($grz$) cutouts of the ten highest \Oion\ mergers, illustrating the impact of redshift on the observed morphology. The figure is organized in two panels (top and bottom), each displaying galaxies ordered by increasing \Oion\ from left to right. Each panel column shows an individual galaxy, identified by its SDSS ObjID at the top. The rows show, from top to bottom, the original image, the image artificially redshifted to $z=0.141$, and to $z=0.26$ (respectively the minimum and average redshifts of Green Peas in \cite{Cardamone2009}). The original redshift, \Oion, merger stage or projected separation to companion are annotated on the original image.} 
    \label{fig:highO32_redshifted}
\end{figure*}

Green Pea and blueberry galaxies are a population of extremely compact star-forming galaxies with strong oxygen emission \citep{Cardamone2009,Yang2017}. Notably, they tend to exhibit high \Oion\ values, and many are known LyC emitters \citep{JaskotOey2013}. What triggers the strong starbursts in this population remains a subject of debate in the literature. On the one hand, most environmental studies find that these galaxies reside in relatively underdense regions (\cite{Cardamone2009,Laufman2022,Kimsey2024,Gupta2026}, although see \cite{Arroyo-Polonio2026}). This has led many authors to argue that merger interactions play a negligible role in triggering the starbursts in these objects. On the other hand, a few 21cm interferometric studies mapping their hydrogen gas have revealed asymmetric and perturbed gas reservoirs indicative of tidal interactions \citep{Purkayastha2022,Purkayastha2024,Dutta2024}. These observations have instead suggested that Green Pea galaxies are the result of galaxy mergers.

The results presented here provide a way to reconcile these two seemingly contradicting views. Rather than being genuinely isolated star-forming galaxies, our findings indicate that some Green Peas with high \Oion\ could possibly be low-mass mergers observed near or post-coalescence. This evolutionary stage naturally explains the lack of close companions in environmental studies, as the galaxies have already merged, or are too close to be resolved. If the progenitor galaxies were of low mass and formed a large fraction of their stars during the merger, older stellar tidal features would be extremely faint, and consequently hard to detect. Yet, because neutral hydrogen is highly sensitive to gravitational interactions \citep{Holwerda2011,Pearson2016}, extended tidal \hi\ features could still remain observable, thus aligning with the findings of interferometric studies.

To test this hypothesis, we have searched for potential overlap between low-redshift Green Pea samples \citep[also referred to as blueberry galaxies;][]{Yang2017} and the general emission line catalog we compiled prior to applying our selection cuts. Unfortunately, due to the inherently low stellar masses of these galaxies ($\rm{M}_{*}\sim10^{7}\,\rm{M_{\odot}}$) and the lack of SDSS spectra for many of them, there is no overlap between our emission line catalog and known local Green Peas. Cross-matching the low-$z$ Green Pea catalog with the Zoobot DESI morphological catalog yields only a single match, with $\texttt{merging\_merger\_fraction} = 0.26$. This value places the galaxy in an ambiguous regime, preventing a definitive classification as either a merger or an isolated system. 

As an additional test, we show the impact of artificially redshifting DESI cutouts for the ten mergers with highest \Oion\ in the Cosmic Disco and pair samples on Figure \ref{fig:highO32_redshifted}. Specifically, we redshift DESI cutouts respectively to the minimum ($z=0.141$) and average ($z=0.26$) redshifts of Green Peas in \cite{Cardamone2009}. The methodology employed for artificially redshifting cutouts is described in Appendix \ref{app:redshifting}. For all but one merger visually identified in Cosmic Disco (ObjID = 1237657630603346040), the low-surface brightness features and double nuclei indicating a merger is taking place have disappeared by $z=0.141$ due to surface brightness dimming and angular size decreasing with redshift. The merger with highest \Oion\ is in a pair with a galaxy outside the cutout ($D_{sep}=20.4\,\rm{kpc}$), and could be identified as a merger if both pair members were observed spectroscopically. The eight remaining mergers appear as compact green sources that could not be accurately identified as mergers from either visual classification or spectroscopic pair search. This supports the hypothesis that some Green Peas could plausibly be late-stage galaxy mergers. Ultimately however, deep, high-resolution optical imaging and \hi\ interferometric observations are required to properly assess the hypothesis that Green Peas are low-mass coalescing galaxy mergers. 

\subsection{Caveats}
While this study marks a step forward in our understanding of how mergers impact ionization properties in galaxies, we note several caveats. Currently, due to instrument limitations, the nature of mergers with high \Oion\ as LyC emitters cannot be evaluated directly in a large enough sample to enable statistically robust analysis. The Habitable Worlds Observatory is a NASA mission currently planned as the next large UV space-based observatory \citep{Dressing2026}. Its throughput in the far-UV will be significantly improved as compared to the Hubble Space Telescope, enabling LyC observations down to $z\sim0.1$ \citep{Citro2025}. It will enable LyC follow-up campaigns of low-z galaxy samples, that will be important to validate the role of galaxy mergers on LyC emission.

Another caveat is due to the apertures used to measure \Oion. In the redshift range considered in this study, the SDSS aperture only samples $0.4-2\,\rm{kpc}$. This represents a small portion of an entire galaxy, and is not necessarily reflective of global \Oion\ values. While the simultaneous matching in redshift and $\rm{M}_{*}$ ensures the scales sampled in mergers and their controls are similar, aperture placement likely leads to some of the \Oion\ scatter observed. Using integral field unit-based observations, such as those delivered by the MaNGA survey \citep{Bundy2015} would help better assess the global \Oion\ properties of mergers and their controls.

Finally and as already noted for pairs, it is difficult to fully characterize the timescale of a merger interaction observationally, and proxies such as projected separation introduce confusion. This also applies to visually identified mergers. Using simulations or simulation-derived timescales would help obtain more precise estimates of the evolution of \Oion\ with merger progression.

\section{Conclusion}\label{sec:conclusion}
We have examined the impact of mergers and merger stages on the ionization state of galaxies via the emission line ratio \Oion. We find a continuous enhancement in \Oion\ as a function of merger progression as traced by the projected pair separation and the visual merger advancement score $s_{merg}$. In particular, we find that mergers have significantly larger median and 90th percentile \Oion\ than isolated galaxies with similar properties past the first pericenter interaction. Coalescing mergers have the largest enhancement in \Oion\, with median $\widetilde{\rm{O}_{32}}=0.92$ and 90th percentile $\rm{O}_{32_{90}}=3.44$. We examine the properties of galaxy mergers with high ionization state ($\rm{O}_{32}>3$), and find they are predominantly galaxy mergers with low stellar masses at late stages of interaction (near coalescence), with significantly larger \Oion\ than their control galaxies. These results confirm previous findings obtained in small LyC-emitting galaxy samples on the role of late-stage galaxy mergers in enhancing the ionization properties of galaxies. Additionally, it demonstrates for the first time that mergers also enhance the ionization state of galaxies with low stellar masses ($\rm{M_{*}<10^9\,\rm{M}_{\odot}}$). Since large values of \Oion\ are a necessary condition for ionizing radiation escape, these results imply galaxy mergers likely facilitate LyC escape from galaxies, in particular at late interaction stages. Finally, we hypothesize that Green Pea galaxies are low-mass mergers at coalescence, which, if verified, would reconcile otherwise contradictory observations of this population. Further studies with next-generation UV telescopes such as the Habitable Worlds Observatory will help confirm the trends identified here through proxy measurements on LyC emission.

\section*{Data Availability}
The data that forms the basis for the analysis presented here are obtained from the SDSS and DESI, and are publicly available. The catalog produced as part of the \textit{Cosmic Disco} Zooniverse project is publicly available via Zenodo (\url{https://doi.org/10.5281/zenodo.23321030}).
 
\begin{acknowledgments}
This study was made possible by the 1099 volunteers who contributed their time to classify galaxies in the Zooniverse project \textit{Cosmic Disco: Characterizing Galaxy Collisions}. This publication uses data generated via the Zooniverse.org platform, development of which is funded by generous support, including a Global Impact Award from Google, and by a grant from the Alfred P. Sloan Foundation.
The authors acknowledge the Minnesota Supercomputing Institute (MSI) at the University of Minnesota for providing resources that contributed to the research results reported within this paper. 
This work made use of the \texttt{legacystamps} package (https://github.com/tikk3r/legacystamps). 
Funding for the Sloan Digital Sky 
Survey IV has been provided by the 
Alfred P. Sloan Foundation, the U.S. 
Department of Energy Office of 
Science, and the Participating 
Institutions. 
SDSS-IV acknowledges support and 
resources from the Center for High 
Performance Computing  at the 
University of Utah. The SDSS 
website is www.sdss4.org.
SDSS-IV is managed by the 
Astrophysical Research Consortium 
for the Participating Institutions 
of the SDSS Collaboration including 
the Brazilian Participation Group, 
the Carnegie Institution for Science, 
Carnegie Mellon University, Center for 
Astrophysics | Harvard \& 
Smithsonian, the Chilean Participation 
Group, the French Participation Group, 
Instituto de Astrof\'isica de 
Canarias, The Johns Hopkins 
University, Kavli Institute for the 
Physics and Mathematics of the 
Universe (IPMU) / University of 
Tokyo, the Korean Participation Group, 
Lawrence Berkeley National Laboratory, 
Leibniz Institut f\"ur Astrophysik 
Potsdam (AIP),  Max-Planck-Institut 
f\"ur Astronomie (MPIA Heidelberg), 
Max-Planck-Institut f\"ur 
Astrophysik (MPA Garching), 
Max-Planck-Institut f\"ur 
Extraterrestrische Physik (MPE), 
National Astronomical Observatories of 
China, New Mexico State University, 
New York University, University of 
Notre Dame, Observat\'ario 
Nacional / MCTI, The Ohio State 
University, Pennsylvania State 
University, Shanghai 
Astronomical Observatory, United 
Kingdom Participation Group, 
Universidad Nacional Aut\'onoma 
de M\'exico, University of Arizona, 
University of Colorado Boulder, 
University of Oxford, University of 
Portsmouth, University of Utah, 
University of Virginia, University 
of Washington, University of 
Wisconsin, Vanderbilt University, 
and Yale University.
We acknowledge the use of Gemini 3.5 Flash and 3.5 Thinking (Google) for language editing and for improving the readability of the manuscript.
The authors thank Claudia Scarlata, Lucy Fortson, Hayley Roberts and Alexander Kuhn for their helpful insights and discussions during various stages of this work.
\end{acknowledgments}

\begin{contribution}
The authors contributed equally to this manuscript.
\end{contribution}

\facilities{Sloan, Bok, Mayall, Blanco}

\software{numpy \citep{numpy}, astropy \citep{astropy1,astropy2,astropy3}, matplotlib \citep{matplotlib}, scipy \citep{scipy}, Jupyter \citep{Granger2021}, astroquery \citep{Ginsburg2019} 
 }

\bibliography{bibliography}{}
\bibliographystyle{aasjournalv7}

\appendix
\section{Comparing merger and control galaxy properties}
\begin{figure*}[h]
    \centering
    \includegraphics[width=0.9\linewidth]{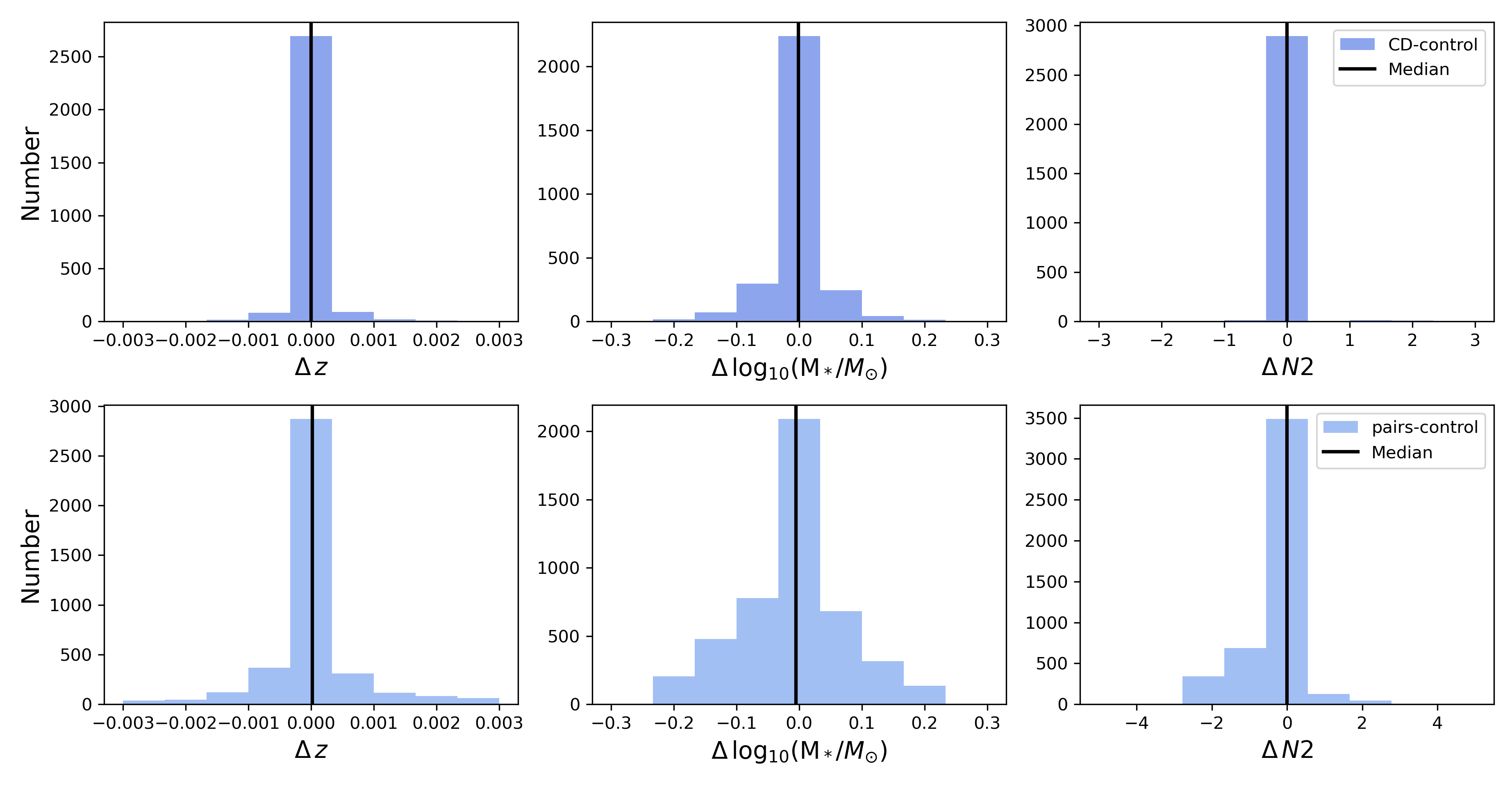}
    \caption{Histograms showing the difference between merger properties and their controls. The top row shows differences in properties between Cosmic Disco mergers and their controls, while the bottom row shows differences in properties between galaxies in pairs and their controls. From left to right, panels present the difference in redshift, stellar mass, and $N_2$.}
    \label{fig:controlvmerg}
\end{figure*}

In Figure \ref{fig:controlvmerg} we show the difference in redshift, stellar mass and $N_2$ for mergers and their controls. All property differences are centered around zero, however, pairs have a $N_2$ difference distribution slightly skewed towards negative values. Thus, while most controls tend to have the same $N_2$ as the mergers, those that don't tend to have slightly larger $N_2$ than pair galaxies. This is due to the way the matching was done for this property, as we required controls for galaxies with large $N_2$ ($>20$) to have $N_2$ within 10\% of that of the merger. Thus, all galaxies with $|\Delta N_2|>1$ are galaxies with $N_{2}>9$. There are few control galaxies with large $N_2$ in the isolated sample (only 2\% of the parent control sample has $N_2>20$), meaning the same control galaxy with large $N_2$ likely served as a control for multiple mergers, resulting in the skewed distribution. Since $N_2$ is not correlated with \Oion, this does not impact the results presented here.

\section{A statistical assessment of the difference in \Oion\ in mergers and their controls}
\begin{figure*}[h]
    \centering
    \includegraphics[width=\textwidth]{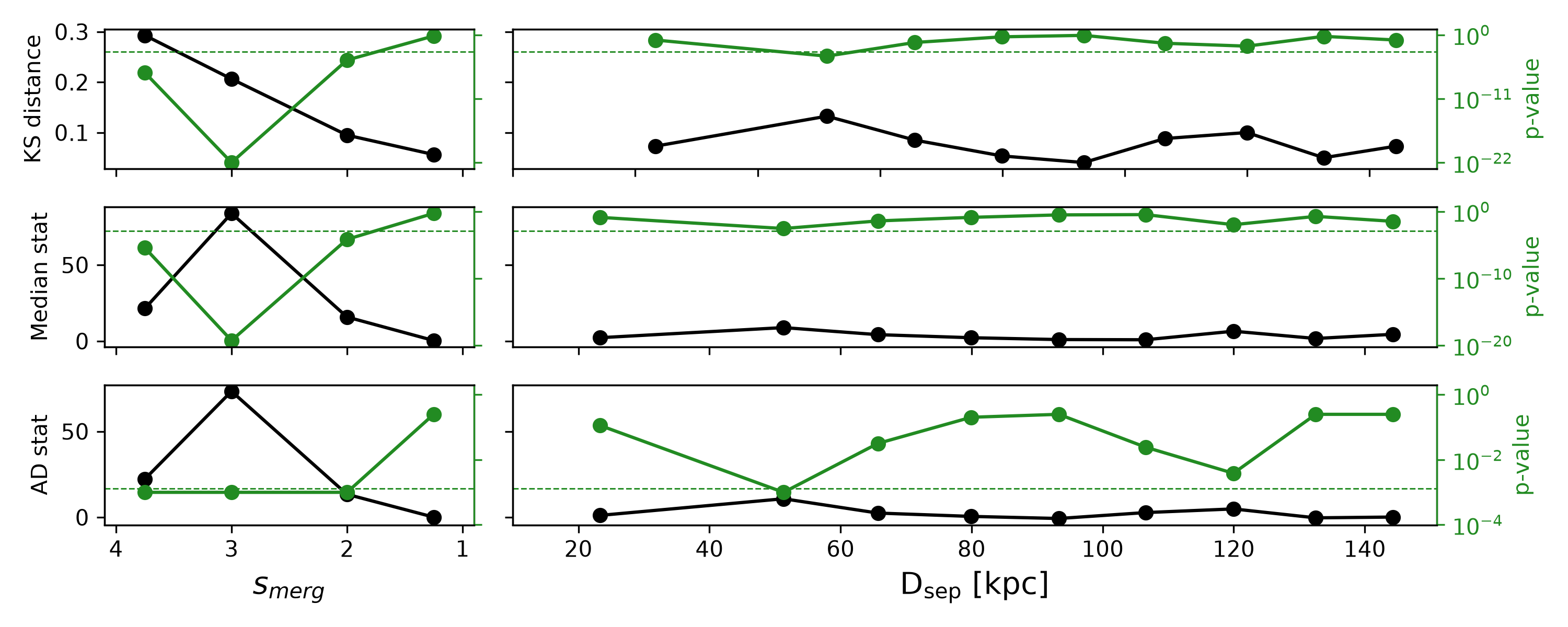}
    \caption{Results of statistical tests ran between galaxy mergers and their control galaxies in the pair sample (right panels) and the Cosmic Disco sample (left panels). The tests are conducted in the same bins as those used in Figure \ref{fig:O32_evol}. From top to bottom, we show results from the Kolmogorov-Smirnov, Mood's median, and Anderson-Darling tests. Black markers and curves show the test statistic values (left axes), while green markers and curves show the evolution of the $p-$value (right axes). The horizontal dashed green lines indicate the threshold for statistical significance ($p<1.3\times10^{-3}$).}
    \label{fig:O32_stats}
\end{figure*}

In Figure \ref{fig:O32_stats} we show the results of statistical tests aiming to determine whether the mergers and their controls have statistically different \Oion\ properties. The tests are done in the same bins as in Figure \ref{fig:O32_evol}. Specifically, we show the results of Kolmogorov-Smirnov (KS) tests, Anderson-Darling (AD) tests and Mood's median tests for mergers and their controls in different bins of merger stage or projected separations. Both the KS and AD tests allow one to test the null hypothesis that two samples are drawn from the same distribution, but the AD test is more sensitive to the tails of a distribution. Mood's median test, on the other hand, tests whether two samples come from distributions with the same median. For most galaxies in pairs, the null hypotheses for these tests cannot be rejected, except in one of the bins ($\sim 50\,\rm{kpc}$), indicating that generally, we cannot reject the null hypothesis that pairs and their controls are drawn from the same distribution. However, the null hypotheses for the KS, AD and Mood's median tests can be rejected with $p<1.3\times10^{-3}$ for mergers in Cosmic Disco with merger stages $s_{merg}>1.5$. This indicates that past the first pericenter interaction, mergers have significantly different \Oion\ properties as compared to control galaxies.

\section{Artificially redshifting DESI images}
\label{app:redshifting}
To artificially redshift DESI cutouts from $z_{obs}$ to $z_{new}$, we account for the evolution of angular size with redshift, cosmological surface brightness dimming, spectral redshifting and match the final images to have the same PSF and noise properties as the original cutout. For each individual frame ($g, r$ and $z$), we apply the following method. First, frames are background-subtracted using \texttt{astropy} sigma-clipped statistics and images are spatially regridded using the angular diameter scale factor $S=D_{a}(z_{obs})/D_{a}(z_{new})$. Pixel flux densities (in nanomaggies) are scaled by the cosmological surface brightness dimming factor $\left(  \frac{1+z_{obs}}{1+z_{new}} \right)^3$, and by a K-correction factor derived from synthetic photometry applied to the SDSS spectrum. Prior to computing the K-correction, the SDSS spectrum is scaled to DESI photometry, corrected for Galactic extinction using the \citep{Cardelli1989} law, and re-extinguished after the spectrum has been shifted to $z_{new}$. The image is then convolved with a Gaussian kernel with $\rm{FWHM}=\sqrt{{(FWHM_{band}\,(1-S))^2}}$, where we assume a FWHM of $1\arcsec.3$, $1\arcsec.2$ and $1\arcsec.1$ respectively for the $g$, $r$ and $z$ bands \citep{Dey2019}. Finally, Gaussian noise is added to the redshifted frames to match the sky background properties of the original image.

\end{document}